\documentclass[twocolumn,epjc3]{svjour3}
\usepackage{amsmath,amsfonts,amssymb}
\usepackage{graphicx}
\usepackage{bm}
\usepackage{braket}
\usepackage{ulem}
\usepackage[colorlinks=true,urlcolor=blue,citecolor=blue,linkcolor=blue]{hyperref}
\usepackage{orcidlink}

\begin{document}

\title{Photon-mediated thermodynamics and density fluctuations in an ensemble of laser-cooled Cesium atoms}

\author{
Arnab Ghosh$^{1,*}$ \and
Inderpreet Kaur$^{1,*}$ \and
Hemant Yadav$^{1}$ \and
Shiva Kant Dwivedi$^{2, 3}$ \and
R Shanmukesh$^{2, 3}$ \and
Mithun Thudiyangal$^{2,3}$ \and
Matthew J. Davis$^{4}$ \and
Bodhaditya Santra$^{1}$
}

\institute{
$^{1}$ Department of Physics, Indian Institute of Technology Delhi, New Delhi 110016, India \\
$^{2}$ Department of Physics and Electronics, Christ University, Bengaluru, Karnataka 560029, India \\
$^{3}$ Center for Quantum Technologies and Complex Systems (CQTCS), Christ University, Bengaluru, Karnataka 560074, India \\
$^{4}$ ARC Centre of Excellence for Engineered Quantum Systems, School of Mathematics and Physics,
University of Queensland, St Lucia, Queensland 4072, Australia \\
\\
$^{*}$ These authors contributed equally to this work. \\
\email{$\mathrm{inderpreetkaur\_90@yahoo.com}$; $\mathrm{bsantra@physics.iitd.ac.in}$}
}

\maketitle

\begin{abstract}
\textcolor{black}{
We present an experimental study of detuning-dependent properties of a laser-cooled cesium cloud in a magneto-optical trap. Fluorescence images are used to extract the cloud size, shot-to-shot width fluctuations, optical depth, density profiles, and spatial density fluctuation spectra as the trapping-laser detuning is varied.  Near resonance, the cloud exhibits larger spatial extent, increased width fluctuations, higher optical depth, and enhanced density-fluctuation power, while larger detunings produce a more reproducible and spatially confined cloud. The measured density profiles are analysed phenomenologically using a generalized Lane--Emden model with a polytropic equation of state, yielding detuning-dependent effective fit parameters in a weak-interaction regime. Power-spectrum and autocorrelation analyses reveal reproducible scale-dependent density correlations. The results provide a quantitative characterization of detuning dependent radiative and collective effects in a cesium MOT and establish a basis for future measurements that can more directly test nonequilibrium transport and photon-mediated interaction models.}
\end{abstract}

\section{Introduction}
{\color{black}Magneto-optical traps (MOTs) provide a robust platform for producing laser-cooled atomic gases~\cite{Phillips1998Nobel,Intro11}, constituting the initial stage toward Bose-Einstein condensation. Although often discussed in terms of single-particle cooling and trapping, a MOT is intrinsically an open, driven-dissipative many-body system. At sufficiently high optical depth, photons scattered by one atom may be reabsorbed by others, leading to radiation trapping, multiple scattering, and effective collective forces that can modify the size, density profile, and stability of the trapped cloud~\cite{weiss2018subradiance,rodrigues2016equation}.  Multiple light scattering fundamentally limits the phase-space density attainable through optical techniques alone~\cite{Intro12}, giving rise to strong Coulomb-like correlations that map onto a one-component trapped plasma~\cite{Intro10}, with quantum corrections to the collective dynamics becoming relevant at typical temperatures of $\sim 100\mu$K. 

Several theoretical and experimental studies have used hydrodynamic or effective thermodynamic descriptions to characterize large laser-cooled clouds. In particular, Rodrigues \textit{et al.} introduced an equation-of-state approach in which the stationary density distribution of a laser-cooled gas is described using a polytropic relation and a generalized Lane--Emden equation~\cite{rodrigues2016equation}. In this framework, the fitted parameters provide an effective description of the balance between trapping, pressure-like terms, and radiative interactions. Such models are useful phenomenological tools, but their application to a MOT requires care because the system is not in global thermodynamic equilibrium. Collective radiative effects in cold atomic gases can also generate spatial and temporal fluctuations of the density. Previous work has discussed radiation trapping, light diffusion, collective modes, and photon-bubble-like instabilities in optically thick atomic clouds~\cite{giampaoli2021photon,romain2016spatial,weiss2018subradiance,IK20,IK23}. }

In this work, we experimentally characterize how the properties of a laser-cooled $^{133}$Cs gas confined in a MOT vary with trapping-laser detuning. We use fluorescence imaging to measure the detuning dependence of the cloud size, shot-to-shot width fluctuations, atom number, optical depth, density profile, and spatial density-fluctuation correlations. We then compare the measured density profiles with numerical solutions of a generalized Lane--Emden model, treating the resulting parameters as effective phenomenological quantities rather than as proof of complete thermodynamic equilibrium. It is useful to distinguish between two types of fluctuations analysed below. First, we study shot-to-shot fluctuations of the global cloud width extracted from Gaussian fits to fluorescence images. These characterize variations in the overall size and reproducibility of the trapped ensemble. Second, we analyse local column-density fluctuations through power-spectrum and autocorrelation methods after removing the mean density profile. These measurements probe spatial correlations in the density field. The results are consistent with detuning-dependent collective radiative effects.

\textcolor{black}{The present work applies the generalized Lane–Emden framework to a complementary experimental regime. Compared with the large 
$^{85}$Rb MOT studied by Rodrigues \textit{et al.}~\cite{rodrigues2016equation}, which contained $10^
7$--$10^{10}$ atoms and operated in the strong multiple-scattering regime, our compact $^{133}$Cs
MOT contains approximately $10^5$ 
 atoms and operates with a total trapping laser power of only 7~mW, corresponding to a comparatively weak multiple-scattering regime. This complementary parameter regime enables us to test the robustness of the generalized Lane--Emden thermodynamic description and to quantitatively compare the extracted effective thermodynamic parameters with those reported by Rodrigues \textit{et al.}\\
The paper is organized as follows. Section~2 describes the experimental setup and the fluorescence image acquisition procedure.  Section~3 investigates the detuning-dependent collective properties of the atomic cloud through measurements of cloud-width fluctuations, mean vibrational occupation number, and optical depth. Section~4 investigates the effective equation of state using the generalized Lane–Emden formalism. Finally, Section~5 analyzes the statistical properties of the density fluctuations using power spectral density and spatial autocorrelation analyses.}

\section {Experimental details} 
\begin{figure}[!t]
    \centering
    \includegraphics[width = 0.52\textwidth]{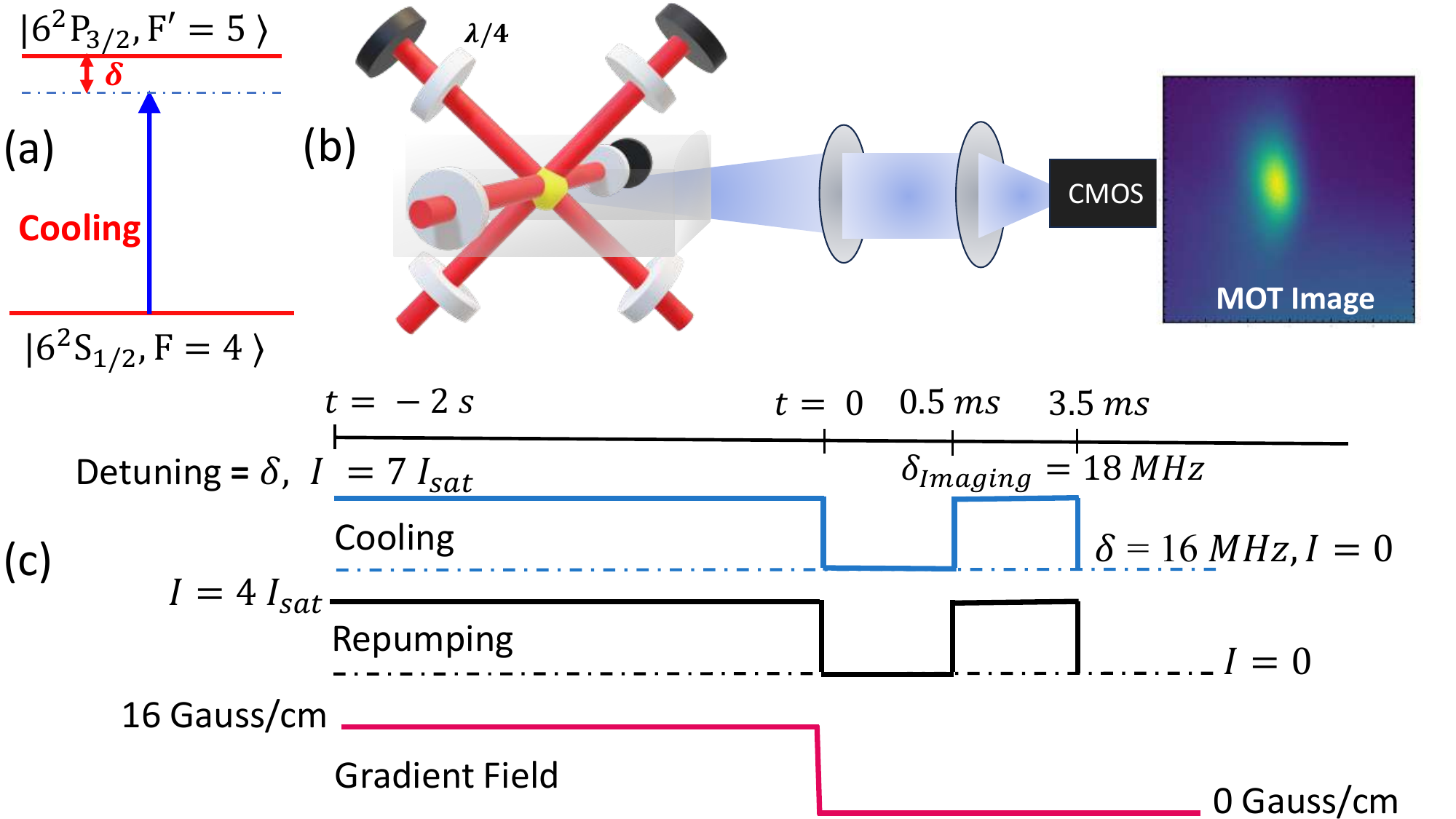}%
   \caption{ \textbf{Experimental setup and timing sequence for 3D MOT of \textsuperscript{133}Cs.}
(a) Energy level diagram showing the cooling and repump transitions used for laser cooling. 
(b) Schematic of the 3D magneto-optical trap formed by three retro-reflected laser beams and a quadrupole magnetic field; fluorescence is collected via a 4f imaging system and detected using a CMOS camera.  Representative MOT fluorescence image is shown on the right. 
(c) Timing diagram illustrating the experimental sequence: the MOT is loaded for 2 seconds with cooling and repump lasers and magnetic gradient field on; at $t = 0$, all fields are turned off allowing atoms to expand freely for 0.5 ms, followed by a 3 ms imaging pulse with cooling and repump beams turned back on for fluorescence collection.}
 \label{02_setup} 
\end{figure}

$^{133}\mathrm{Cs}$ atoms are collected from a dilute atomic vapor at a background pressure of $\sim10^{-9}$ mbar using a three-dimensional magneto-optical trap. The trapping laser operates on the $D_2$ line of cesium at 852 nm and is frequency-locked to the transition between the ground state (F = 4) and the excited state (F$'$ = 5). The trapping beam of 8 mm $1/e^2$-diameter and a total power of 7 mW, is red-detuned by $\delta = 18 $ MHz from resonance. The double-pass acousto-optic modulator (AOM) is aligned in parallel with the spectroscopy~\cite{chen2019single,yadav2025optimization} setup to enable precise tuning of the locked frequency without altering the optical power.
Additionally, the MOT's magnetic field gradient is set to 16 Gauss/cm, and the repump beam is slightly detuned (by 2 MHz) from the F = 3 to F$'$ = 4 transition. Under these conditions, we obtain a cold atomic cloud containing approximately $10^5$ atoms, with a spatial extent of 0.5 mm $1/e^2$-diameter. The temperature of the cloud is around 100 $\pm$ 5~$\mu$K. The experimental setup is shown in Fig.~\ref{02_setup}.
\par
After loading the MOT for 2 s, all trapping and repump beams, along with the magnetic field, are turned off. Following a 500~$\mu$s free fall, a fluorescence image of the atomic cloud is captured using a CMOS camera with a 3 ms exposure time, providing the integrated spatial distribution along the z-axis. During the MOT loading phase, the laser detuning ($\delta$) is set close to resonance to enhance atom capture through strong photon scattering and to enable exploration of regime dominated by multiple scattering in dense atomic clouds. For imaging, the detuning is increased to 18 MHz  to suppress the effect of multiple scattering, thereby enabling precise spatial characterization of the atomic distribution under controlled conditions.
\par 

To explore the cloud fluctuations emerging from photon–atom interactions and their effect on the cloud characteristics, we vary the red detuning of the trapping laser in the range $\delta = 4~\mathrm{MHz}$ to $\delta = 14~\mathrm{MHz}$. For each detuning value, we capture and average 30 fluorescence images, taken with both the trapping and repump beams active, to minimize statistical fluctuations. This range allows us to examine how variations in detuning influence the properties of the cold atomic cloud and provides insight into the onset of turbulence. 

For the statistical analysis presented in Fig.~\ref{hist}, we acquire a larger dataset of 500 fluorescence images for selected detuning values, ensuring improved visualization and reliable extraction of the cloud width distributions. 

The images used for spatial characterization are acquired at $\delta = 18~\mathrm{MHz}$, which helps reduce distortions arising from scattered light and interatomic interactions. This larger detuning improves measurement accuracy by minimizing reabsorption effects, enabling a clearer determination of the atomic cloud shape for different MOT loading detunings.

\section {\color{black}Detuning-dependent collective effects in a cold atomic gas }
\textcolor{black}{Near resonance, enhanced photon scattering increases the radiation pressure, leading to an expansion of the MOT cloud and an increase in atom number. To characterize the resulting collective behavior, we first analyze the shot-to-shot variations in the overall cloud size obtained from repeated fluorescence images. Throughout this work, these variations are referred to as \textit{cloud-width fluctuations}. We further investigate the mean vibrational occupation number and the optical depth to characterize the collective properties of the atomic cloud. \\
\\
The reabsorption of scattered photons gives rise to photon-mediated interactions that can influence the collective behavior of the atomic cloud. Accordingly, the measured cloud-width fluctuations, mean vibrational occupation number, and optical depth provide indirect signatures of these collective effects. As the detuning is increased away from resonance, the atomic scattering rate decreases, weakening multiple scattering and photon-mediated interactions. Consequently, the cloud becomes more stable and tightly confined, exhibiting reduced cloud-width fluctuations, lower optical depth, and lower temperatures due to the reduced momentum transfer from scattered photons.
}
\subsection{Statistical analysis of cloud width fluctuations}\label{CW1}
\begin{figure}[b!]
    \centering
\includegraphics[width=.49\textwidth]{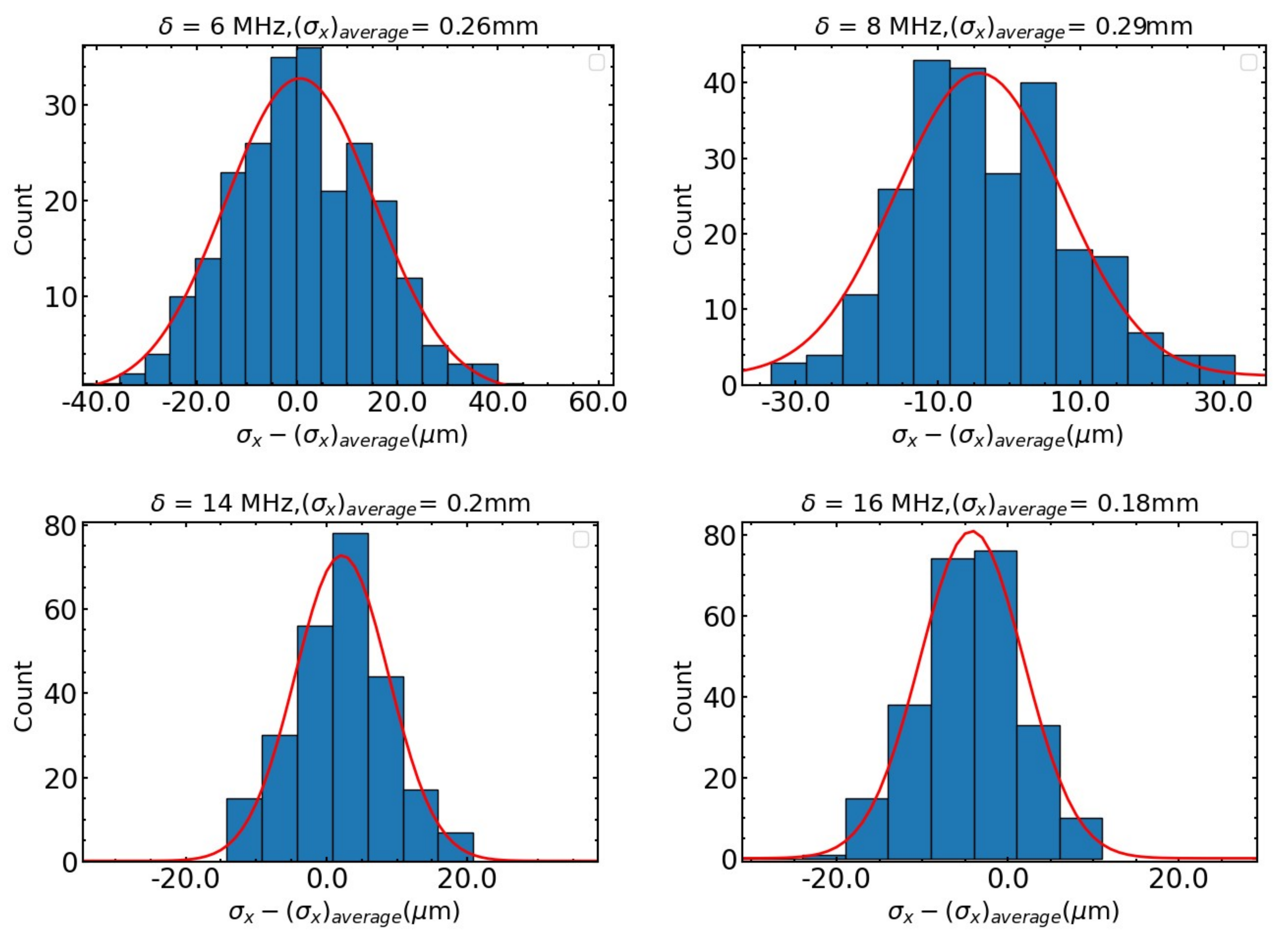}
    \caption{\color{black}
Histograms of cloud-width fluctuations for $\delta = 6$, 8, 14, and 16 MHz, obtained from 500 fluorescence images. The horizontal axis represents the deviation from the mean width, $\sigma_x-\langle\sigma_x\rangle$. The red curves are Gaussian fits used to determine the fluctuation spread.}\label{hist}
\end{figure}

We collected 500 fluorescence images of the cold-atom cloud for each detuning value
($\delta = 6, 8, 14,$ and $16~\mathrm{MHz}$) to obtain statistically reliable information about the
cloud's behavior. For each image, the central region of the cloud was isolated and a
one-dimensional density profile was extracted from the fluorescence intensity. These
profiles were fitted with Gaussian functions to determine the cloud width $\sigma$ for
individual realizations. Repeating this procedure for all images yielded distributions of
$\sigma$, which were compiled into histograms for each detuning as shown in Fig.~\ref{hist} and subsequently fitted
with Gaussian functions to extract the mean cloud width and the corresponding spread. 

The histogram analysis reveals a clear detuning-dependent evolution of the cloud size
and stability, characterized by the histogram-averaged width $\sigma_{\mathrm{histogram}}$.
At lower detunings, broad histograms indicate enhanced shot-to-shot fluctuations in the
cloud width. Specifically, the fitted mean values (Fig.~\ref{hist}) were
$\sigma_{\mathrm{histogram}} = 14~\mu\mathrm{m}$ at $\delta = 6~\mathrm{MHz}$ and
$\sigma_{\mathrm{histogram}} = 11~\mu\mathrm{m}$ at $\delta = 8~\mathrm{MHz}$, reflecting
stronger density variations and reduced stability. As the detuning increases, the
histograms become progressively narrower, indicating improved reproducibility of the
cloud size. For $\delta = 14~\mathrm{MHz}$ and $\delta = 16~\mathrm{MHz}$, the extracted
widths decrease to $\sigma_{\mathrm{histogram}} = 6~\mu\mathrm{m}$ and
$\sigma_{\mathrm{histogram}} = 5~\mu\mathrm{m}$, respectively, demonstrating a transition
toward a more stable and tightly confined cloud configuration.

This behavior can be understood from the detuning dependence of the atomic scattering
cross section, which governs the strength of light--atom interactions. For near-resonant
light the scattering cross section follows
\begin{equation}
\sigma(\Delta) \propto \frac{1}{1 + 4\Delta^2/\Gamma^2},
\end{equation}
where $\Delta$ denotes the detuning from resonance and $\Gamma$ is the natural linewidth of the transition. Increasing the detuning therefore reduces the effective scattering probability, weakening multiple scattering and photon reabsorption processes inside the atomic ensemble. As a result, density fluctuations are suppressed and the cloud becomes more stable at larger detunings.

\subsection{Mean vibrational occupation number}\label{widthcloud}
For each detuning value, Gaussian fits to the one-dimensional fluorescence profiles extracted from individual images yield a set of cloud widths $\{\sigma_i\}$, reflecting shot-to-shot fluctuations of the atomic ensemble. The spread of these fitted widths, defined as the standard deviation $\sigma_{\mathrm{spread}}=\mathrm{Std}(\sigma_i)$, provides a statistically robust measure of the spatial extent of the cloud.

{\color{black}Assuming harmonic confinement, one can define a quantity \cite{srivathsan2019measuring} 
\begin{equation}
\bar{n} = \frac{m\omega\sigma_{\mathrm{spread}}^{2}}{\hbar}-\frac{1}{2}, 
\label{eq: mean occupation number}
\end{equation} 
where $\bar{n}$ is the mean occupation number and provides a quantitative measure of the motional excitation of the trapped atomic ensemble. Larger values of $\bar{n}$ correspond to stronger motional excitation, whereas smaller values indicate tighter confinement and lower motional energy.
}

{\color{black}
In eq. \eqref{eq: mean occupation number}, $m$ is the mass of a cesium atom, $\omega$ is the effective trap frequency ($\omega/2\pi \approx 469\,\mathrm{Hz}$ for our MOT), and $\hbar$ is the reduced Planck constant. Eq  \eqref{eq: mean occupation number} suggests a quadratic relation between $\bar{n}$ and $\sigma_{spread}$. $\bar{n}$ scales quadratically with $\sigma_{\mathrm{spread}}$, rendering it an especially sensitive indicator of motional excitation: small decreases in the spread are amplified into significantly larger reductions in $\bar{n}$. This quadratic sensitivity makes $\bar{n}$ a powerful figure of merit for assessing the degree of thermal excitation in the atomic cloud. Here, the analysis is performed using a data set of 30 realizations for each detuning value.}

\begin{figure}[t!]
    \centering
\includegraphics[width=.49\textwidth]{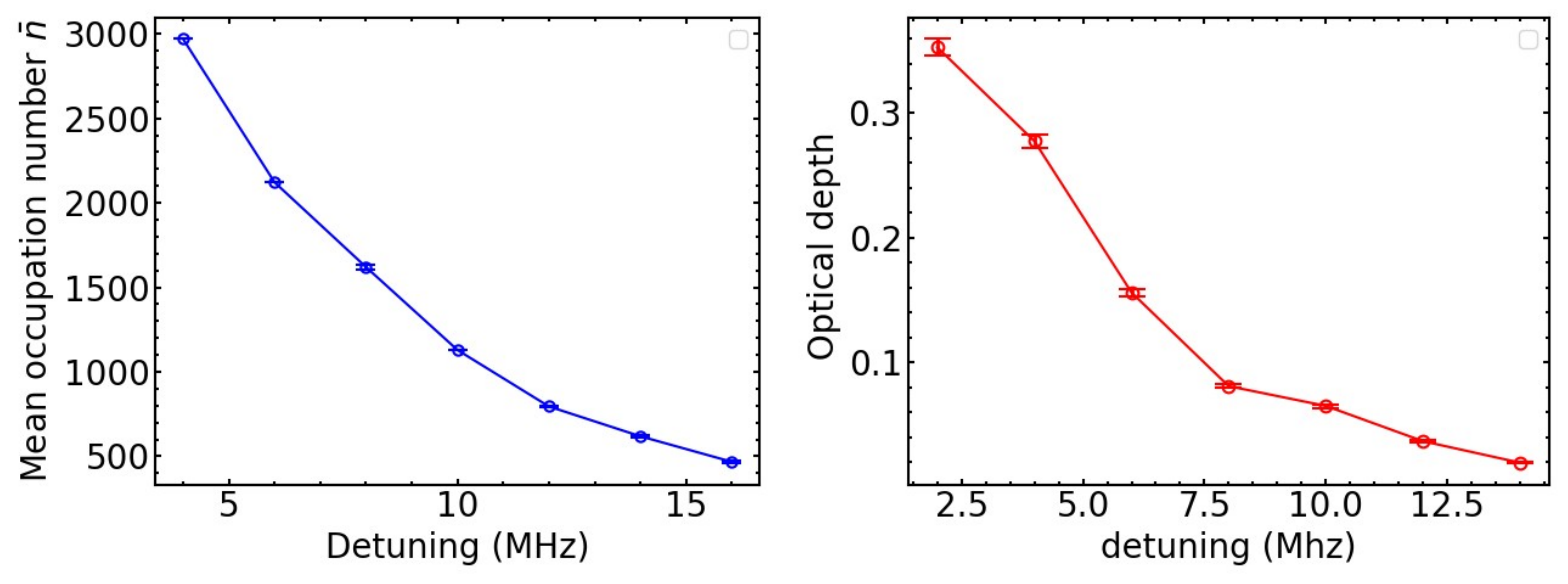}
    \caption{{\color{black}(a) Mean vibrational occupation number, $\bar{n}$, calculated from the measured spread of the cloud-width distributions using Eq.~(2).  (b) Optical depth determined from the measured atom number and cloud size using Eq.~(3). with error bars reflect the propagated \textcolor{black}{1$\sigma$} uncertainty from the Gaussian fit to the cloud radius. }}\label{n_bar}
\end{figure}

The experimentally determined spreads, ranging from $\sigma_{\mathrm{spread}} = 1.20 \times 10^{-4}\,\mathrm{m}$ at $\delta = 4~\mathrm{MHz}$ to $3.54 \times 10^{-5}\,\mathrm{m}$ at $\delta = 16~\mathrm{MHz}$, yield a corresponding sequence of $\bar{n}$ values that decrease steadily with increasing detuning. Specifically, $\bar{n}$ drops from approximately $2.97 \times 10^{3}$ at $\delta = 4 ~\mathrm{MHz}$ all the way to $4.67 \times 10^{2}$ at the largest detuning of $\delta = 16~\mathrm{MHz}$, as seen in Fig.~\ref{n_bar}(a).

The monotonic suppression of $\bar{n}$ across this detuning range is a direct
signature of enhanced spatial confinement and diminished thermal fluctuations
within the ensemble. Taken together, these results demonstrate that the
detuning-dependent evolution of $\bar{n}$ faithfully captures the progressive
reduction in motional energy, {\color{black} suggesting an increasing stabilization of the atomic cloud with detuning.}

\subsection{Optical depth }\label{OD}
\textcolor{black}{The optical depth is a key parameter for characterizing dense cold-atom ensembles, as it quantifies the strength of light scattering and photon-mediated multiple scattering within the MOT. Its dependence on laser detuning provides direct insight into the evolution of radiation trapping and collective optical effects. To determine the optical depth, both the atom number and the cloud size are extracted from the recorded fluorescence images.}\\\\
\textcolor{black}{ Contrary to the method in section \ref{CW1} where we analyze variation of $\sigma$ in each image for different detuning, here for each detuning, we extract the effective radius from \textit{averaged} fluorescence images.}
The effective cloud radius $R$ is obtained by integrating the recorded fluorescence signal along one spatial axis and fitting the resulting one-dimensional density profile with a Gaussian function, where the fitted width $\sigma$ is taken as the cloud radius. \textcolor{black}{The fluorescence counts within the cloud region are then converted into the corresponding number of detected photons by accounting for the camera quantum efficiency, exposure time, and the collection solid angle of the imaging system.} The atomic scattering rate is evaluated using the standard two-level atom model, incorporating the effects of the saturation parameter, the natural linewidth $\Gamma$, and the laser \textcolor{black}{detuning $\delta$}. \textcolor{black}{The atom number $N$ is then obtained by comparing the detected photon flux with the calculated scattering rate while accounting for the overall detection efficiency of the imaging system.}

The optical depth $b$ is subsequently calculated using the relation~\cite{weiss2018subradiance}
\begin{equation}
b = g \, \frac{3N}{(kR)^2} \, \frac{1}{1 + 4(\delta/\Gamma)^2},
\end{equation}
where $g$ denotes the degeneracy factor of the Cs D$_2$ transition and $k$ is the optical wavevector. \textcolor{black}{For the Cs D$_2$ transition, $g=7/15$, $k=2\pi/\lambda$ with $\lambda=852.357\,\mathrm{nm}$, and $\Gamma=5.2\,\mathrm{MHz}$.} This analysis is repeated for multiple detuning values, and the resulting optical depth is plotted as a function of detuning, as shown in Fig.~\ref{n_bar}(b). A systematic decrease in optical depth with increasing detuning is observed, reflecting the reduction of the {atomic scattering cross}\cite{weiss2018subradiance} section away from resonance, which weakens radiation trapping and photon-mediated multiple-scattering processes, thereby reducing the effective optical density of the atomic cloud.
 
\par

\section{Effective equation of state of the cold atomic gas}

An equation of state relates thermodynamic variables, such as pressure and density, and provides an effective macroscopic description of a many-body system. In astrophysics, the Lane--Emden equation is widely used to describe self-gravitating polytropic fluids, where the pressure and density are related through a polytropic equation of state,
\begin{equation}
P=K\rho^{1+\frac{1}{\gamma}},
\end{equation}
where $P$ and $\rho$ denote the pressure and density, respectively, $K$ is a proportionality constant, and $\gamma$ is the polytropic index in the stellar equation of state~\cite{Cox1968}.

Inspired by this approach, an effective thermodynamic description has been developed for laser-cooled atomic gases. In the absence of a complete microscopic theory for a MOT, the effective pressure is assumed to obey a polytropic equation of state \cite{PhysRevA.88.023412},
\begin{equation}
P=C_\gamma n^\gamma,
\end{equation}
where $P$ is the effective pressure of the atomic gas, $n$ is the atomic density, $C_\gamma$ is a constant related to the thermodynamic properties of the system, and $\gamma$ is the effective polytropic exponent. The value of $\gamma$ reflects the balance between thermal motion and photon-mediated collective interactions.

Rodrigues \textit{et al.}~\cite{rodrigues2016equation} adapted the
Lane--Emden formalism by introducing an effective equation-of-state
description for a laser-cooled gas in a MOT. Although a MOT is
intrinsically a driven-dissipative system, the generalized Lane--Emden
equation is employed here as an effective phenomenological description
of the measured density profiles, following the framework of Rodrigues
\textit{et al.}~\cite{PhysRevA.88.023412,rodrigues2016collective}. The resulting equation is
\begin{equation}
\frac{1}{\zeta^2}\frac{d}{d\zeta}\left(\zeta^2\frac{d\theta}{d\zeta}\right)
+ \Omega \, \theta^{\gamma-1} = 0 ,
\label{eq: lane_emden_equation}
\end{equation}

where $\theta(\zeta)$ is the normalized density, $\zeta$ is the dimensionless radial coordinate, and $\Omega$ represents the strength of the effective long-range interaction arising from multiple photon scattering. This formalism is described in detail by Rodrigues \textit{et al.}~\cite{rodrigues2016equation} and is briefly summarized in ~\ref{Appsec: Lane emden formalism}.

Experimentally, the measurements are performed after a 2 s loading period, during which the MOT reaches a stationary state. The atomic density is
obtained from the fluorescence image by extracting a one-dimensional
cross-sectional line profile through the cloud center, as shown by the
red line in Fig.~\ref{04_analysis}. The profile is obtained by
averaging over 30 recorded fluorescence images and is first fitted with
a Gaussian function to provide an empirical description of the measured
density distribution.

The generalized Lane--Emden equation is then solved
numerically over a range of the fitting parameters
$(\gamma,\Omega)$ to generate model density profiles. For each parameter
pair, the calculated density profile is interpolated onto the
experimental spatial coordinates and compared with the experimentally
measured density profile using a least-squares metric. The optimal values
of $\gamma$ and $\Omega$ are obtained by minimizing this least-squares
error over the explored parameter space. The corresponding best-fit numerical solution is then
compared directly with both the experimentally measured
density profile and the Gaussian fit (~\ref{ADSC},  Fig.~ \ref{fig:lane_emden_comparison}). The
parameter $\gamma$ characterizes the effective equation of state of the
atomic gas, while $\Omega$ quantifies the strength of the effective
photon-mediated interaction within the phenomenological model.

\begin{figure}[t!]
    \centering
\includegraphics[width=.49\textwidth]{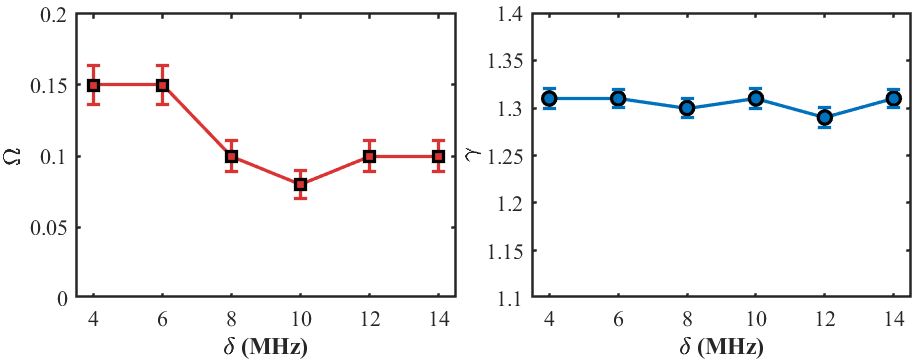}
    \caption{\textcolor{black}{Variation of the interaction parameter $\Omega$ (left) and the
polytropic exponent $\gamma$ (right) with trapping-laser detuning.
The parameters are extracted by fitting numerical solutions of the
generalized Lane--Emden equation to the experimentally measured
density profiles. The error bars represent the estimated uncertainties
associated with the fitted parameters.}}
    \label{Variation}
\end{figure}

The resulting dependence of $\gamma$ and $\Omega$ on the laser detuning is presented in Fig.~\ref{Variation}. We observe that the normalized interaction parameter $\Omega$ increases as the detuning approaches resonance, suggesting an increased contribution from photon-mediated interactions  and enhanced collective effects in this regime. This behaviour reflects the increasing role of multiple scattering processes in determining the effective thermodynamic properties of the laser-cooled atomic cloud. 
The extracted values of the polytropic exponent $\gamma$ and the interaction parameter $\Omega$ may be compared with those reported by Rodrigues \textit{et al.}~\cite{rodrigues2016equation}. While the values of $\gamma$ obtained in the present work remain within the range reported in Ref.~\cite{rodrigues2016equation}, the interaction parameter $\Omega$ is considerably smaller, reflecting the comparatively weaker multiple-scattering regime explored here. This difference is consistent with the substantially lower atom number ($\sim10^5$ atoms), reduced optical depth, and lower trapping laser power employed in the present experiment.
Correspondingly, unlike the large MOT studied by Rodrigues
\textit{et al.} \cite{rodrigues2016equation}, where the density distribution
evolved toward a flat-top profile, the measured density
profiles in our experiment remain well described by
Gaussian functions. A representative comparison between the measured profile, the Gaussian fit, and the corresponding best-fit numerical Lane--Emden solution is presented in  ~\ref{ADSC}.

\par

\section {Power spectrum analysis of the density fluctuations}
\textcolor{black}{Unlike the cloud-width fluctuations discussed in Sec.~\ref{CW1}, which describe shot-to-shot variations in the overall cloud size, the density fluctuations analyzed here correspond to local spatial variations in the atomic density within individual fluorescence images. We analyze the experimental results using the power spectral density (PSD) and the  radial autocorrelation function of the atomic density.}
The power spectral density quantifies the amplitude of spatial density fluctuations. To obtain the PSD, we acquire up to 30 fluorescence images of the atom cloud under the previously described experimental conditions. These images provide the spatial atom distribution integrated along the line of sight, denoted as $n_i(\mathbf{r})$, where $\mathbf{r} = (x,y)$ and $i = 1, 2, \dots, 30$.

We compute the PSD of column-density fluctuations by first defining the normalized density $\tilde{\rho}_i(\mathbf{r}) \equiv n_i(\mathbf{r}) / n_0(\mathbf{r})$, where $n_0(\mathbf{r}) = \langle n_i(\mathbf{r}) \rangle$ is the ensemble-averaged density profile. The corresponding fluctuation field is $\delta \tilde{\rho}_i(\mathbf{r}) \equiv \tilde{\rho}_i(\mathbf{r}) - \langle \tilde{\rho}_i(\mathbf{r}) \rangle$. The PSD is then obtained as $|F_i(\mathbf{k})|^2$, where $F_i(\mathbf{k})$ is the two-dimensional Fourier transform of $\delta \tilde{\rho}_i(\mathbf{r})$. The ensemble-averaged PSD, $E(\mathbf{k}) = \langle |F_i(\mathbf{k})|^2 \rangle$~\cite{giampaoli2021photon,Steinhauer2017}, provides the spectral distribution of density fluctuations. Assuming isotropic fluctuations, we further perform an angular average to obtain the radially averaged PSD, $E(k)$.

\begin{figure}
    \centering \includegraphics[width=\linewidth]{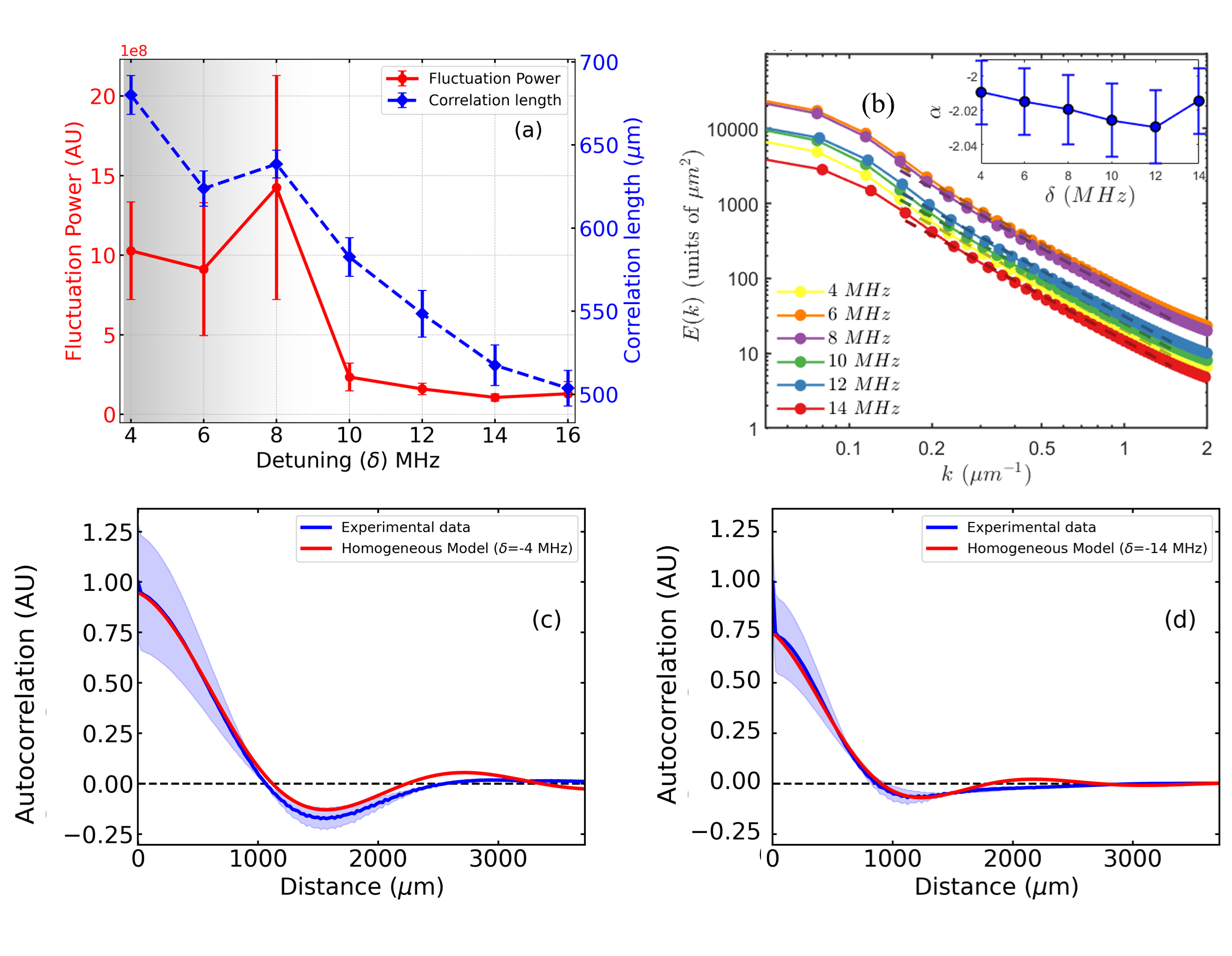}
        \caption{ (a) Correlation length and total power of relative atom density fluctuations as a function of laser detuning. The observed trends  \textcolor{black}{are consistent with a gradual evolution toward a more fluctuation-dominated regime. The error bars correspond to {\color{black}$1\sigma_{SD}$} deviation from the mean across $30$ different realizations.} (b) Power spectral density, $E(k)$, for various values of the detuning parameter $\delta$. The inset shows the scaling exponent $\alpha$ obtained for different values of $\delta$. (c) Radial autocorrelation function (in blue) in the \textcolor{black}{near resonance} regime and (d) \textcolor{black}{far from resonance} regime. \textcolor{black}{The shaded region represents $1\sigma$ deviation from the mean across $30$ different realizations. The red curve represents the fit to the spherically symmetric solutions of the homogeneous photon bubble model on the experimental data.}}
    \label{fig:pf and cl}
\end{figure}

Figure~\ref{fig:pf and cl}(a) shows the total power of the atom density fluctuations—obtained by integrating the power spectral density (PSD) over all accessible wavenumbers—together with the corresponding correlation length (cf. Sec.~\ref{Auto correlation}). Around \(\delta = 8\) MHz, the fluctuation amplitude rises sharply, accompanied by a roughly linear increase in the correlation length. \textcolor{black}{ This detuning-dependent evolution of the correlation length is consistent with evolution of $\bar{n}$ and optical depth in Fig. \ref{n_bar} and again suggest the increasing stabilization of atomic cloud with increasing detuning. Also, it further can be interpret as the detuning-dependent increase in amplitude and spatial range of density fluctuations}.

\textcolor{black}{
Figure~\ref{fig:pf and cl}(b) shows the PSD calculated from density profiles measured at detuning values between 4 and 14~MHz. In the log-log representation, the PSD exhibits an approximate power-law decay, $E(k)\propto k^{-\alpha}$, over the fitted range. Linear fits yield exponents close to $\alpha \simeq 2$, with high coefficients of determination. This reproducibility indicates robust scale-dependent structure in the measured density fluctuations.
}

\par

We emphasize that $E(k)$ is a PSD of column-density fluctuations, rather than a kinetic-energy spectrum. The effective steepening may reflect system-specific factors such as compressibility, dimensionality reduction due to the imaging geometry, finite-size effects, or external confinement. The proximity to a $k^{-2}$ dependence is consistent with scenarios where density fluctuations inherit features of compressible or two-dimensional cascades. Importantly, the reproducibility of the exponent across different detuning values suggests a robust underlying mechanism governing the fluctuation dynamics in the atom cloud.

Panels (c) and (d) present the experimentally measured radial autocorrelation function alongside the homogeneous model prediction from \cite{giampaoli2021photon}, for detunings in the \textcolor{black}{near and far resonant regimes, respectively. The error bars correspond to $1\sigma$ variations over $30$ different realizations of the experiment for each detuning.} The model captures the behavior well at small to intermediate scales \textcolor{black}{(\(R^2 \geq 0.98\) and stays within the statistical limits. In the near resonance case ($\delta = 4$MHz) the homogeneous model deviates vastly at large length scales indicating the breakdown of the homogeneous approximation and reflects the importance of finite-size effects as long range fluctuations develop and the system enters into an unstable regime.}

     \section{Discussions and conclusions } 
    
    {\color{black}We experimentally investigated the detuning-dependent behaviour of a laser-cooled cesium cloud in a magneto-optical trap. As the trapping laser is tuned closer to resonance, the cloud exhibits increased spatial extent, atom number, optical depth, and shot-to-shot width fluctuations, whereas larger detunings produce a more compact and reproducible cloud. The measured density profiles are well described by a generalized Lane--Emden model with an effective polytropic equation of state, whose fitted parameters vary systematically with detuning. Owing to the driven-dissipative nature of a MOT, these parameters are interpreted as effective steady-state descriptors rather than evidence of global thermodynamic equilibrium. Power-spectrum and autocorrelation analyses reveal detuning-dependent spatial correlations and finite correlation lengths, indicating that the cloud properties evolve with changes in scattering rate and optical depth. Although these observations are consistent with photon-mediated collective effects, they do not uniquely establish the underlying physical mechanism. Overall, the present results provide a quantitative experimental characterization of the influence of laser detuning on the density distribution, optical depth, cloud-width fluctuations, and spatial correlations, providing a reliable experimental foundation for future investigations of collective radiative effects in laser-cooled atomic gases.
}

\section*{Acknowledgments}
MT acknowledges the support of Anusandhan National Research Foundation (ANRF), Government of India, through the prime minister early career grant ANRF/ECRG/2024/003150/PMS, and Christ University for funding the research through the seed grant, Sanction No. CU-ORS-SM-24/94. The authors acknowledge stimulating discussions with Poonam Yadav.
\bibliographystyle{unsrt}
\bibliography{ref}
\appendix
\clearpage

\section{Supplemental Materials}
\subsection{Lane-Emden Formalism} \label{Appsec: Lane emden formalism}
We briefly summarize the hydrodynamic framework used to model the system, following Ref.~\cite{rodrigues2016equation}. A fluid description of a laser-cooled gas confined in a
magneto-optical trap (MOT) may be introduced with the usual continuity and Navier-Stokes equations\cite{mendonca2013physics}\cite{Collectiveoscillationsinanultracoldatomicgas}\cite{PhysRevA.88.023412}:
\begin{equation}
   \frac{\partial n}{\partial t} + \nabla \cdot (n v) = 0,
   \label{diffusion equation}
\end{equation}
\begin{equation}
    \frac{\partial v}{\partial t} + v \cdot \nabla v = -\frac{\nabla P}{mn} + \frac{F_t}{m} + \frac{F_c}{m},\label{momEq}
\end{equation}
where \(n\), \(v\), and \(m\) denote density, velocity, and atomic mass, respectively.
In this framework, the multiple scattering force $F_c$ is described by:
\begin{equation}
   \nabla \cdot F_c = Q n,
\end{equation}
with $Q$ denoting the effective charge of the atoms. In the momentum equation (\ref{momEq}), \(F_t\) represents the trapping force of the magneto-optical trap, which acts as a harmonic damping force on the atomic cloud due to the anti-Helmholtz coil.

At this point, to explore the thermodynamic properties of the system, we  consider the polytropic equation of state, which relates pressure to density as\cite{PhysRevA.88.023412}:
\begin{equation}
   P(r) = C_\gamma n(r)^\gamma,
\end{equation}
where \(C_\gamma\) is a constant and \(\gamma\) is the polytropic index. By substituting this relation into the hydrodynamic equations, we arrive at a generalized equation of state in the form of the Lane-Emden equation:
\begin{equation}
   \frac{1}{\zeta^2} \frac{d}{d\zeta} \left( \zeta^2 \theta^{\gamma - 2} \frac{d\theta}{d\zeta} \right) - \Omega \theta + 1 = 0,
\end{equation}
where the density is expressed as \(n(r) = n(0) \theta(r)\), and the radial coordinate \(r\) is scaled by \(a_\gamma = \sqrt{\frac{C_\gamma}{3 m \omega_0^2}} n(0)^{(\gamma - 1)/2}\), with \(\zeta = r / a_\gamma\).

The equation is solved numerically to obtain density profiles for given \(\Omega\) and  \(\gamma\). Here,  \(\Omega\) controls the transition to the multiple-scattering regime, while  \(\gamma\) characterizes deviations from ideal gas behavior.
The parameter \(\Omega\) significantly influences the cloud distribution, with \(\Omega = 0\) corresponding to a stable cloud and \(\Omega = 1\) indicating the onset of the multiple-scattering regime. Meanwhile, \(\gamma\) characterizes deviations from ideal gas behavior; \(\gamma = 1\) describes an ideal gas, while values of \(\gamma\) different from unity reflect instability due to multiple scattering effects in the atomic cloud. 
\subsection{Autocorrelation Analysis of Density Fluctuations}
\label{Auto correlation}

Following Ref.~\cite{giampaoli2021photon}, we compute the mean atomic density over \(N\) experimental realizations. Denote by \(n_i(x,y)\) the measured column density in the \(i\)-th realization, with \(i = 1,2,\dots,N\) (in our case \(N=30\)). The ensemble-averaged density is
\[
\langle n(x,y) \rangle \;=\; \frac{1}{N} \sum_{i=1}^{N} n_i(x,y).
\]
To focus on fluctuations around this average while removing shot-to-shot variations in total atom number, we define for each realization
\[
\delta n_i(x,y) \;=\; n_i(x,y) \;-\; \alpha_i \, \langle n(x,y)\rangle,
\]
where the normalization coefficient \(\alpha_i\) accounts for the total-atom-number difference in the \(i\)-th shot:
\[
\alpha_i \;=\; \frac{\sum_{x,y} n_i(x,y)}{\sum_{x,y} \langle n(x,y)\rangle}.
\]
Here \(\sum_{x,y}\) denotes summation over all camera pixels. This ensures removal of global atom-number fluctuations i.e., \(\sum_{x,y} \delta n_i(x,y) = 0\) for each realization, isolating spatial fluctuations independently of global atom-number drift.

Next, we compute the two-dimensional autocorrelation for each realization:
\[
C_i(\mathbf{r'}) \;=\; \bigl\langle \delta n_i(\mathbf{r}) \,\delta n_i(\mathbf{r} + \mathbf{r'}) \bigr\rangle_{\mathbf{r}} \;=\; \frac{1}{N_{\rm pix}} \sum_{\mathbf{r}} \delta n_i(\mathbf{r}) \,\delta n_i(\mathbf{r} + \mathbf{r'}),
\]
where \(\mathbf{r}=(x,y)\), \(\mathbf{r'}=(\Delta x,\Delta y)\), and the average \(\langle \cdot \rangle_{\mathbf{r}}\) indicates summation (or discrete averaging) over all pixel positions \(\mathbf{r}\) for which \(\mathbf{r} + \mathbf{r'}\) remains within the image bounds. We then average over the \(N\) realizations:
\[
C(\mathbf{r'}) \;=\; \frac{1}{N} \sum_{i=1}^{N} C_i(\mathbf{r'}).
\]
Assuming isotropy, we perform radial averaging to obtain \(c(r)\), with \(r = \|\mathbf{r'}\|\). In practice, one bins values of \(C(\mathbf{r'})\) according to the radius \(r\) and computes the average within each bin:
\[
c(r) \;=\; \langle C(\mathbf{r'}) \rangle_{\|\mathbf{r'}\| = r}.
\]
This radial autocorrelation \(c(r)\) characterizes how density fluctuations at one point correlate with those at distance \(r\). Strong spatial correlations at certain scales may signal underlying structures such as photon-bubble instabilities and associated turbulence regimes.

We define a characteristic correlation length \(L\) from \(c(r)\) by a weighted average:
\[
L \;=\; \frac{\sum_{k} |c(r_k)\,r_k|}{\sum_{k} |c(r_k)|},
\]
where \(\{r_k\}\) are the discrete radii used in the radial average. This definition captures the typical scale over which correlations remain significant.

\paragraph*{Relation to the Power Spectrum.}
The autocorrelation function \(C(\mathbf{r'})\) and the power spectral density \(E(\mathbf{k})\) form a Fourier transform pair:
\[
E(\mathbf{k}) \;\propto\; \mathcal{F}\bigl[C(\mathbf{r'})\bigr](\mathbf{k}), 
\quad
C(\mathbf{r'}) \;\propto\; \mathcal{F}^{-1}\bigl[E(\mathbf{k})\bigr](\mathbf{r'}).
\]
Experimentally, we observe a power spectrum of the form
\[
E(k) \;\propto\; \frac{1}{(a^2 + k^2)^2},
\]
where \(k = \|\mathbf{k}\|\), and \(a>0\) is a small-\(k\) regularization parameter related to the inverse of a characteristic length scale. The inverse Fourier transform of this Lorentzian-squared form yields an exponentially decaying autocorrelation in real space:
\[
c(r) \;\propto\; e^{-a r},
\]
indicating a characteristic correlation length \(\sim 1/a\), associated with photon-bubble structures. This supports the presence of spatial structures with size \(1/a\) in the atomic cloud. An exponential decay in \(c(r)\) confirms the presence of coherent features (e.g., photon bubbles) with that characteristic size.

\subsection{Experimental Analysis} \label{ADSC}
\par

The atom number is estimated by integrating the fluorescence signal. The signal is calibrated using the camera’s exposure time, quantum efficiency, and the effective solid angle of the imaging system, determining the fraction of fluorescence collected.{ As shown in Fig. 2(d), the measured atom number varies slightly between runs. The error bars represent the standard deviation of the total pixel count from multiple measurements, reflecting this variation}.
\par
Assuming a 3D Gaussian profile, the cloud volume is 
\[
V = (2\pi)^{3/2} \sigma_x \sigma_y \sigma_z,
\]
where \(\sigma_x\), \(\sigma_y\), and \(\sigma_z\) are the 1/e\(^2\) radii extracted from Gaussian fits to the fluorescence image. In the case of two-dimensional imaging, \(\sigma_z\) is inferred from geometric symmetry. The atomic density \(n\) is then computed as
\[
n = \frac{N}{V},
\]
where \(N\) is the total atom number extracted from the integrated fluorescence signal. Error bars on the density are derived by propagating uncertainties in both atom number and spatial width measurements.

\begin{figure}[!h]
    \centering
    \includegraphics[width = 0.5\textwidth]{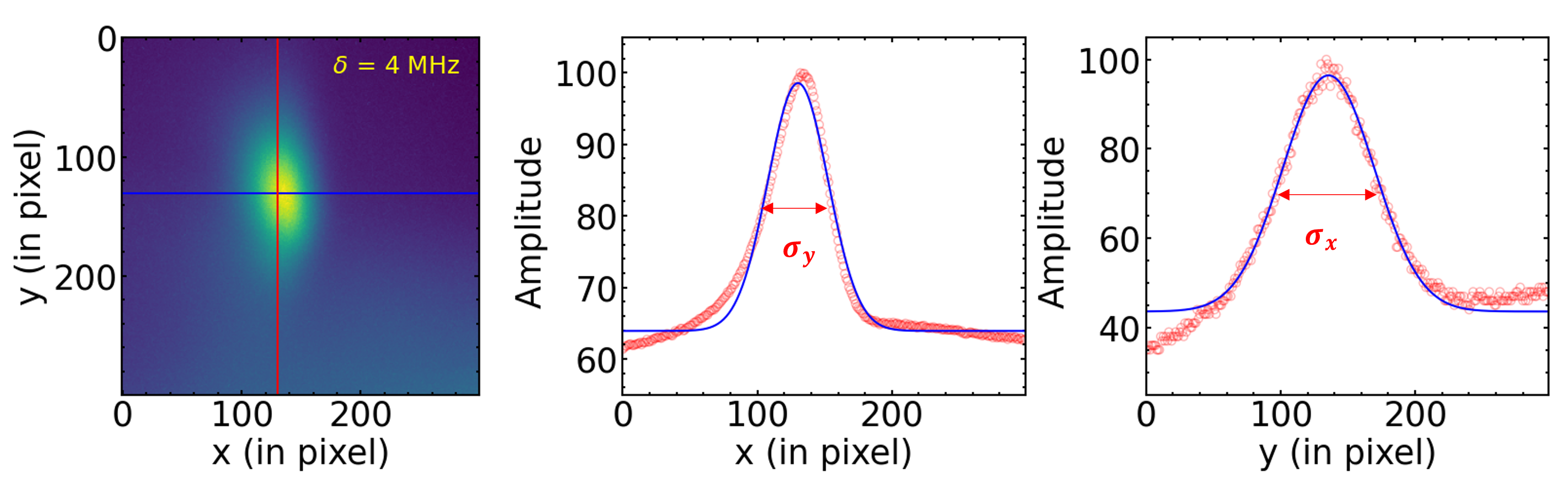}
   \caption{ \small {Equation of the state} : Atomic density of the cloud (MOT fluorescence image) captured by sCMOS camera. The Red line corresponds to the line profile of MOT intensity along the vertical axis. We have fitted Gaussian to this profile and extract the $ \sigma_x $ from the Fit. Similary we have plot blue curve and extract $\sigma_y $ { where Background signal arising from the intrinsic noise of the camera.}}
 \label{04_analysis} 
\end{figure}

The experimental cMOS data are averaged over the independent realizations to improve the signal-to-noise ratio. One-dimensional (1D) profiles are extracted by taking linear cuts through the center of the cloud, defined as the center of mass of the two-dimensional image. In Fig.~\ref{04_analysis}, the blue curves represent these line profiles along the \(x\)- and \(y\)-axes intersecting the center of mass. The profiles are fitted with Gaussian functions (depicted in red), from which the characteristic widths \(\sigma_x\) and \(\sigma_y\) are determined.

\begin{figure}[t]
    \centering
\includegraphics[width=0.5\textwidth]{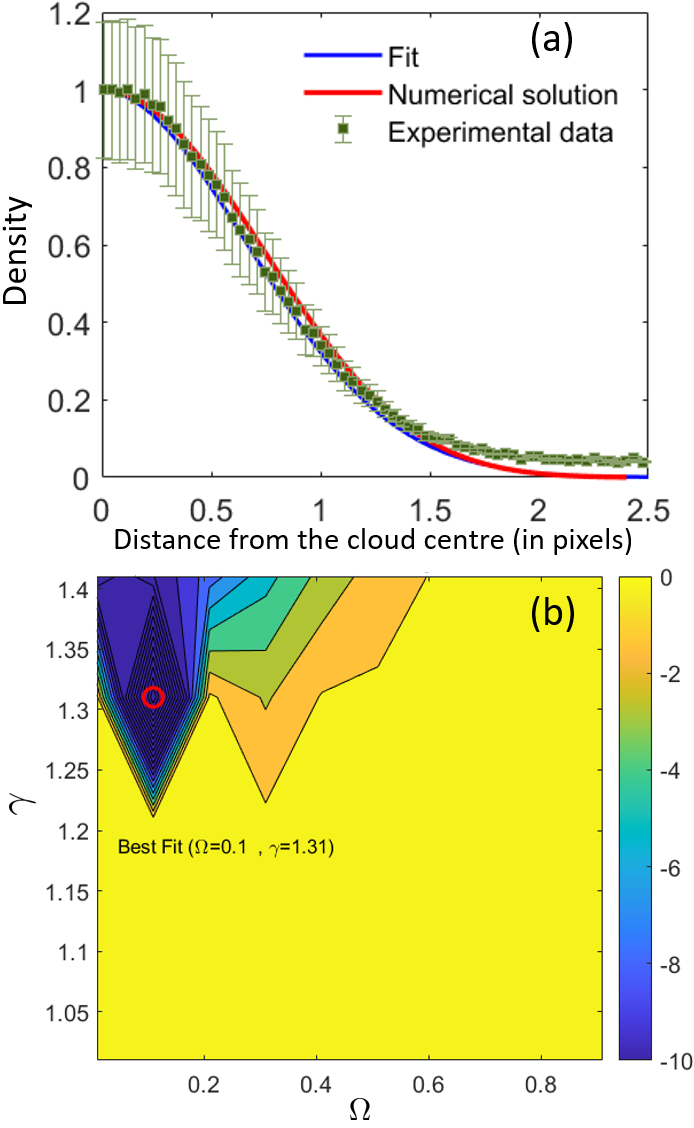}
    \caption{
(a) Comparison of the experimentally measured one-dimensional
density profile (green symbols with error bars), the empirical Gaussian
fit (blue solid line), and the best-fit numerical solution of the
generalized Lane--Emden equation (red solid line) for a trapping-laser
detuning of $\delta = 14$ MHz. (b) Least-squares error landscape for the
same detuning in the $(\Omega,\gamma)$ parameter space. The red circle
indicates the optimal parameter pair,
$(\Omega,\gamma)=(0.10,1.31)$, and the color scale represents
$\log_{10}$ of the least-squares error.
}
    \label{fig:lane_emden_comparison}
\end{figure}

\textcolor{black}{To determine the effective parameters $\gamma$ and
$\Omega$, the generalized Lane--Emden equation is solved
numerically over a two-dimensional grid of parameter
values. The numerical density profile for each
$(\gamma,\Omega)$ pair is interpolated onto the
experimental coordinate grid and compared with the
measured one-dimensional density profile using a
least-squares error metric. The parameter pair
corresponding to the minimum least-squares error is
identified as the optimal solution and used to generate
the numerical profile shown in
Fig.~\ref{fig:lane_emden_comparison}(a). The least-squares error shown in Fig.~\ref{fig:lane_emden_comparison}(b),  exhibits a well-defined minimum in the explored
parameter space. The contours are noticeably narrower along the
$\gamma$ direction than along the $\Omega$ direction, indicating that
the fitting procedure is more sensitive to variations in the polytropic
exponent than in the interaction parameter.\\
Fig.~\ref{fig:lane_emden_comparison}(a) compares the
experimentally measured one-dimensional density profile
(red symbols with error bars) with the empirical
Gaussian fit (green solid line) and the best-fit
numerical solution of the generalized Lane--Emden
equation (blue dashed line) for $\delta=14$ MHz. The
numerical solution closely follows the measured density
profile and remains nearly Gaussian in shape,
demonstrating that the generalized Lane--Emden model
provides an effective phenomenological description of
the experimental density distribution in the weak
multiple-scattering regime investigated here.}

\begin{figure*}[t]
    \centering
    \includegraphics[width = 0.9\textwidth]{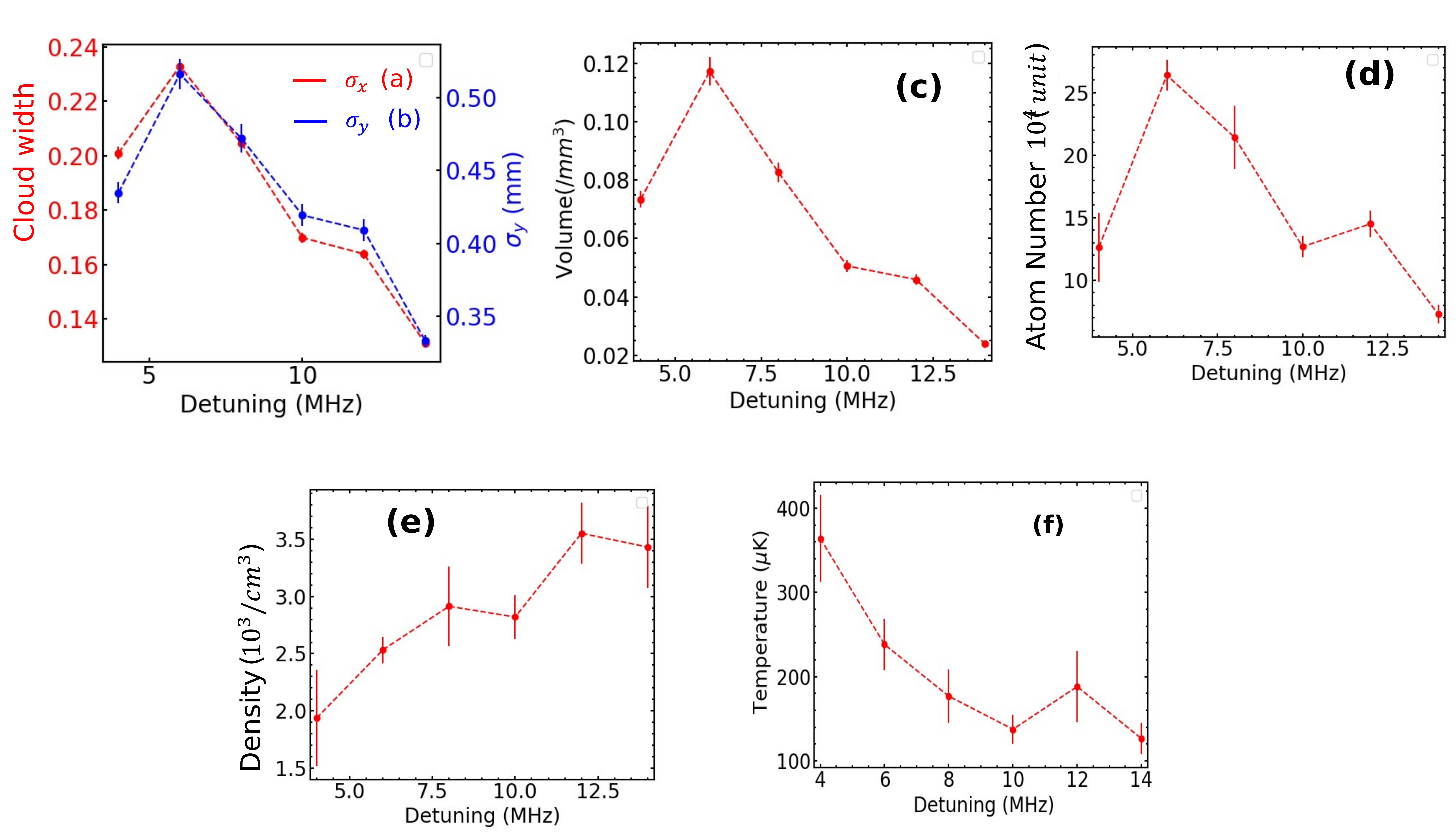}
   \caption{\textbf{Variation of MOT cloud parameters with detuning of the cooling laser}: (a) Cloud's width along x-direction ($\sigma_x$) and (b) along y-direction ($\sigma_y$), (c) cloud volume, (d) Atom number, (e) Atomic density, and (f) Measured temperature using TOF method.
   }
 \label{05_width_analysis} 
\end{figure*}
\subsection {Thermodynamic parameter of cold atomic cloud }
For a fixed magnetic field gradient of $\nabla$B $\sim$ 14 Gauss/cm, we studied the variation in the spatial extent and atom number of a cold atomic cloud as a function of the laser detuning. Our analysis reveals that as the detuning approaches resonance, both the width and volume of the atomic cloud increase significantly, consistent with enhanced photon scattering near resonance. This behavior is illustrated in Fig.~\ref{05_width_analysis}(a)-(c), where the fitted Gaussian widths \(\sigma_x\) and \(\sigma_y\), and the corresponding cloud volume are shown to broaden as the laser detuning decreases toward zero. The full width at half maximum (FWHM) increases due to increased radiation pressure, leading to stronger expansion of the atomic cloud in all spatial directions. 

\par

In Fig.~\ref{05_width_analysis} (a) we have plotted the $\sigma_x $ and (b) $\sigma_y $ with detuning where the error bar corresponds to the fitted error. As the detuning approached resonance, the FWHM increased, which aligns with expectations. This occurs because the atomic cloud expands due to enhanced photon scattering when the laser is tuned closer to resonance. Similar effects have been reported in previous studies, such as Rodrigues et al.~\cite{rodrigues2016collective}, which observed a flat-top profile in the density distribution near resonance, indicating collective effects~\cite{giampaoli2021photon,PhysRevLett.108.033001} and saturation of the imaging beam. However, in our setup, due to the relatively low power of the trapping laser (approximately 7 mW), we did not observe this flat-top feature. Instead, we measured a more gradual increase in the cloud's FWHM, as shown in Fig.~\ref{05_width_analysis}(a),(b).
\par
A similar effect is observed in the volume of the atomic cloud. As the detuning approaches resonance, the increased photon scattering leads to greater expansion, not only in terms of the FWHM but also in the overall volume occupied by the atoms. This expansion occurs because atoms experience stronger radiation pressure when the laser frequency is closer to resonance, causing a more pronounced spread in all spatial directions. Consequently, the cloud's volume increases, reflecting the same underlying mechanism that drives the broadening of its spatial profile as shown in \ref{05_width_analysis}(c).

The number of atoms varies with the detuning of the trapping laser. To account for this effect, we have measured the atom number at different detunings to ensure accurate analysis, Fig.~\ref{05_width_analysis}(d). By tracking these variations, we can properly interpret the changes in density and distinguish them from effects arising purely due to atom loss or redistribution. This allows for a more reliable density analysis by normalizing the data and observing trends that are intrinsic to the system's behavior rather than artifacts of changing atom numbers.
\par
We analyzed the density variations to investigate how the atomic density changes with detuning, as shown in Fig.~\ref{05_width_analysis}(e). As expected, moving closer to resonance enhances the effects of multiple scattering. This is due to two main factors: first, the cross-sections (\(\sigma_R\) and \(\sigma_L\)) increase, leading to stronger interactions between atoms and scattered photons; second, the number of trapped atoms also rises, further increasing the probability that a photon will be reabsorbed before leaving the system. These combined effects contribute to a more significant redistribution of light within the cloud, influencing its overall density profile. \\
The temperature of the atomic cloud was measured at different detunings using the time-of-flight method, as shown in Fig.~\ref{05_width_analysis}(f).  To minimize the effects of multiple scattering, these measurements were conducted with a fixed detuning of 18 MHz. Our results indicate that the temperature decreases as the detuning moves further from resonance. Specifically, near resonance, the temperature is approximately 250~\(\mu\)K, whereas at a detuning of 14 MHz, it drops to around 100~\(\mu\)K. This behavior is expected, as atoms experience stronger photon scattering near resonance, leading to increased momentum transfer and heating. In contrast, at larger detunings, reduced scattering results in lower temperatures.

\end{document}